\documentclass[twocolumn,showpacs,preprintnumbers,amsmath,amssymb,prl]{revtex4}

\usepackage{graphicx}
\usepackage{dcolumn}
\usepackage{bm}

\begin{document}


\title{Anisotropic 2D diffusive expansion of ultra-cold atoms in a disordered potential}
\author{M.~Robert-de-Saint-Vincent, J.-P.~Brantut$^*$, B.~Allard, T.~Plisson, L.~Pezz\'e,
L.~Sanchez-Palencia, A.~Aspect, T.~Bourdel$^\dag$, and P.~Bouyer}
\affiliation{Laboratoire Charles Fabry de l'Institut d'Optique,
Univ Paris Sud, CNRS, campus polytechnique RD128, 91127 Palaiseau
France}

\date{\today}

\begin{abstract}
We study the horizontal expansion of vertically confined
ultra-cold atoms in the presence of disorder. Vertical confinement
allows us to realize a situation with a few coupled harmonic
oscillator quantum states. The disordered potential is created by
an optical speckle at an angle of 30$^{\circ}$ with respect to the
horizontal plane, resulting in an effective anisotropy of the
correlation lengths of a factor of 2 in that plane. We observe
diffusion leading to non-Gaussian density profiles. Diffusion
coefficients, extracted from the experimental results, show
anisotropy and strong energy dependence, in agreement with
numerical calculations.
\end{abstract}

\maketitle

Transport in most materials is determined by the complex interplay
of many ingredients, for instance the structure and thermal
fluctuations of the substrate~\cite{ashcroft1976},
the interparticle interactions, which can induce
superconductivity~\cite{degennes1966} or metal-insulator
transitions~\cite{Mott68}, and disorder~\cite{Anderson58}.
Disorder is relevant to
many condensed-matter systems and strongly affects transport via
scattering. Its primary effect is thus diffusion, an effect
underlying the Drude theory of conductivity~\cite{ashcroft1976},
as well as the self-consistent theory of Anderson
localization~\cite{vollhardt1980}. Disorder is of
special interest in dimension two (2D), which is the marginal
dimension for return probability in Brownian motion and for
Anderson localization \cite{Abrahams79}.
Moreover, intriguing effects, which
are not fully understood, occur in 2D, such as the
metal-insulator transitions in high-mobility Si
MOSFETs~\cite{Abrahams01, Kravchenko04}, GaAs
heterostructures~\cite{Allison06, Tracy09}, and thin metal-alloy
films~\cite{Dubi07}.

Ultra-cold atomic gases are good candidates to study classical or
quantum disordered systems (see
Refs.~\cite{Fallani08,Sanchez-Palencia09} and references therein).
They offer unique versatility as one can control the amount and
type of disorder, the interaction strength or the confinement
geometry. In 1D, Anderson localization~\cite{Billy08,Roati08} and
interaction-induced delocalization~\cite{Deissler09} have been
observed. In 3D, the competition between interaction and disorder
has been investigated in disordered optical
lattices~\cite{White09,Pasienski09}. Diffusion was reported for
speckle-induced 3D optical molasses in the dissipative
regime~\cite{Grynberg00}. So far, less work has been devoted to
2D.

In this letter, we study diffusion of ultra-cold atoms in an
effectively anisotropic disordered potential without dissipation.
The geometry is planar as the atoms are confined vertically to a
size of about $1\,\mu$m  in a dipole trap and horizontally free to
move over a millimeter. In the presence of disorder, we observe
expansion at a reduced speed and anisotropic, non-Gaussian atomic
density profiles. We show that the dynamics is horizontally
diffusive. Fitting a diffusive model to the data, we extract the
diffusion coefficients and find that they are anisotropic and
strongly energy-dependent. Our results are consistent with
numerical simulations assuming classical dynamics.

\begin{figure}[t!]
\centering
\includegraphics[width=0.49\textwidth]{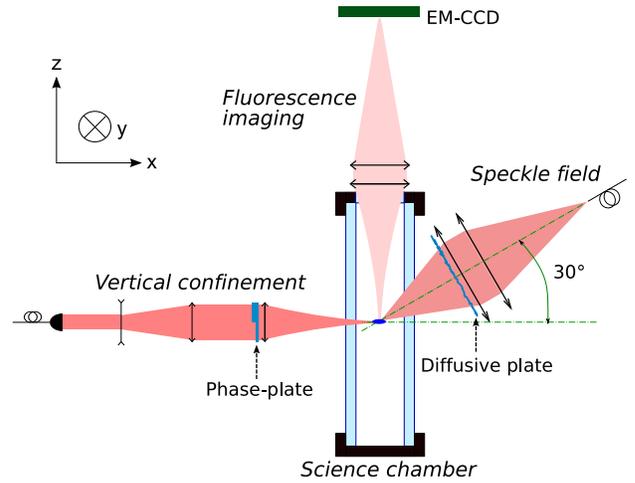}
\caption{Experimental setup. The ultra-cold atom cloud is released
in a vertically confining potential created by a blue-detuned
Hermite-Gauss TEM01-like beam \cite{Meyrath05, Smith05},
prepared with a 0-$\pi$ phase plate (left). The disorder is
induced by an angled speckle light field created from a diffusive
plate (right). Finally, the horizontal cloud expansion is detected
from the top by fluorescence imaging on an Andor Electron
Multiplying Charge Coupled Device (EM-CCD).
\label{fig1}}
\end{figure}
The experimental setup uses a vertically confining potential and a
speckle light field (see Fig.\,\ref{fig1}). Both are created with
767\,nm laser light \cite{Stern09}, blue detuned from the
resonance at 780\,nm for $^{87}$Rb atoms in their ground state.
They thus induce a repulsive potential. The vertical confinement
is realized between the two lobes of a vertically focused
Hermite-Gauss TEM01-like mode, prepared with a holographic 0-$\pi$
phase plate \cite{Meyrath05, Smith05}. The measured vertical
trapping frequency is $\omega/ 2 \pi = 680$\,Hz with a 22\,$\mu$m
separation between the two intensity maxima, a total power of
150\,mW, and an horizontal waist radius at $1/e^2$ of 1.1\,mm.

The speckle light field is produced by a beam passing through a
diffusive plate \cite{Goodman, Clement06} and focused on the
atoms. The high numerical aperture (a 75\,mm diameter aperture at
a distance of 150\,mm from the atoms) allows us to achieve an
approximately Gaussian correlation function with transverse
correlation length $\sigma_y = 0.8 \,\mu$m (half-width at
$1/\sqrt{e}$). The longitudinal correlation length (in the
direction of propagation) is deduced to be $\sigma_\textrm{long}
\approx 9\, \mu$m \cite{Clement06}. As the speckle beam is at
$30^\circ$ from the horizontal expansion plane, the correlation
length $\sigma_x$ along $x$ is $2 \sigma_y=1.6\,\mu$m. With a
power of 66\,mW and a Gaussian waist radius of 1.1\,mm, the
average disordered repulsive potential at the center of the beam
is $\bar V \approx k_{\textrm{B}} \times 53(8)$\,nK\,$\approx (2
\pi \hbar)\times 1.1(2)\,$kHz (where $k_{\textrm{B}}$ is the
Boltzmann constant and $2\pi \hbar$ the Planck constant).

The experiment proceeds as follows. An ultra-cold atom sample is
produced by an all-optical runaway evaporation in a crossed dipole
trap at 1565\,nm, as described in Ref.~\cite{Clement09}. The atom
cloud is first transferred in 5 ms in a trap combining
simultaneously the initial crossed dipole trap and the vertically
confining beam. The crossed trap is then further ramped down in
200\,ms in order to reduce the confinement and thus also the
temperature to $k_{\textrm{B}} T \approx k_{\textrm{B}} \times
200(20)$\,nK\,$\approx (2 \pi \hbar)\times 4.2(4)\,$kHz, slightly
above the condensation threshold. Finally, the speckle field is
ramped up in 4\,ms, and 1\,ms later a thermal cloud of
$N=1.5\times 10^5$ atoms is released in the horizontal plane by
suddenly turning off the crossed dipole trap. After a chosen 2D
expansion time, the vertical confinement and the speckle potential
are switched off. After 0.1\,ms, the atomic column density is
measured from the top through fluorescence imaging.

\begin{figure}[t!]
\centering
\includegraphics[width=0.49\textwidth]{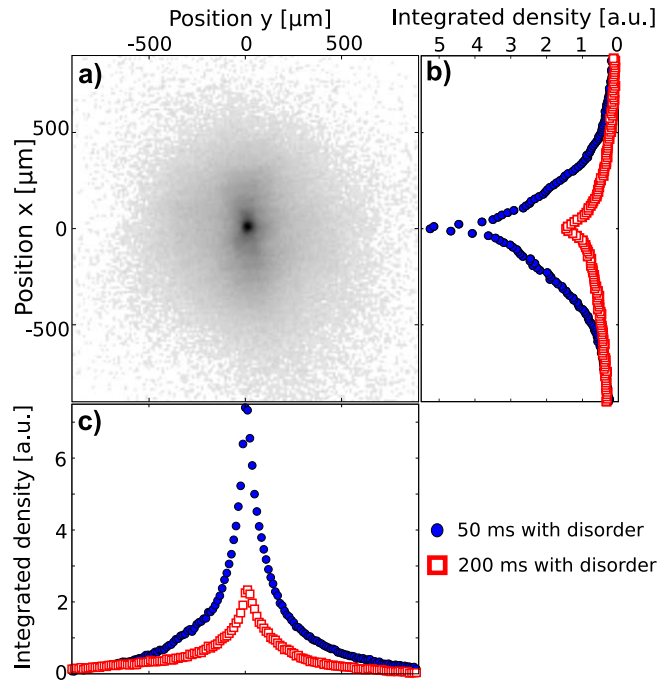}
\caption{Atomic column density after planar expansion of an
ultra-cold gas in an anisotropic speckle potential. a: Image after
50\,ms of expansion. b,c: Integrated density along the two major
axes. The plain dots (open squares) correspond to 50\,ms (200\,ms)
of expansion. \label{fig2}}
\end{figure}
A typical image for an expansion time of 50\,ms in the disordered
potential is presented in Fig.\,\ref{fig2}a. The corresponding
integrated density along $y$ (resp.\ $x$) is plotted in
Fig.\,\ref{fig2}b (resp.\,\ref{fig2}c). We observe a sharp
anisotropic structure elongated along $x$ around the initial
position, surrounded by a broader isotropic cloud similar to what
is observed in the absence of disorder. The sharp anisotropic
structure corresponds to low energy atoms, whose expansion has
been slowed down by the disorder, whereas the broad cloud
corresponds to atoms which expand almost ballistically at this
time scale. For an expansion time of 200\,ms (see Fig.\,2b and
2c), the contribution of the ballistic atoms is negligible with
respect to the lowest energy atoms. The cloud profiles are then
found to be non-Gaussian with long tails in both directions.
Similar profiles have been theoretically predicted for energy
dependent diffusive behavior in the expansion of Bose-Einstein
condensates \cite{Shapiro07}.

\begin{figure}[t!]
\centering
\includegraphics[width=0.4\textwidth]{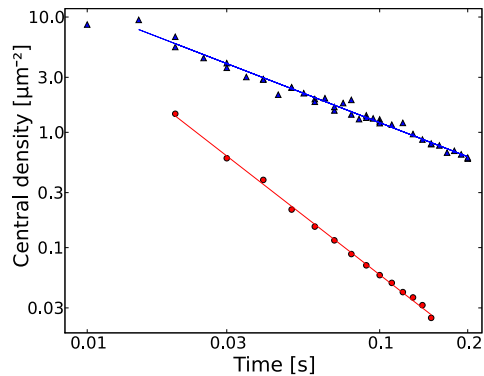}
\caption{Evolution of the peak column density $n(0,0, t)$ as a
function of expansion time. Triangles: with disorder; circles:
without disorder. The solid lines correspond to fits with
algebraic time dependence between 15 and 200\,ms. The fitted
slopes of the decay are -0.98 with disorder and -1.97 without
disorder. \label{fig3}}
\end{figure}
We first study the behavior of the peak column density $n(0, 0,
t)$ as a function of time. It should scale as $1/t$ in a diffusive
regime, and as $1/t^2$ for a ballistic expansion.
Figure~\ref{fig3} shows a log-log plot of the measured peak
density as a function of time, with and without disorder. For
times below 15\,ms, the cloud is smaller than the pixel size and
therefore our measurement does not reflect $n(0, 0, t)$. Between
15 and 200\,ms, we observe a linear behavior with a slope
$-2.0^{-0.2}_{+0.3}$ without disorder, whereas with disorder, we
find a linear behavior with a slope $-1.0^{-0.1}_{+0.3}$. The
uncertainties come from the dispersion of the slopes found for
different data sets taken in similar conditions. This measurement
is consistent with diffusive expansion of a significant part of
the atoms in the horizontal plane. After only 15\,ms, the
contribution of the ballistically expanding atoms to the density
at the origin vanishes.

\begin{figure}[!t]
\begin{center}
\includegraphics[width=0.49\textwidth]{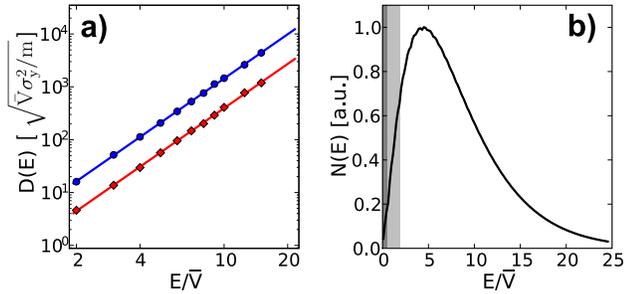}
\end{center}
\vspace{-0.4cm} \caption{a: Diffusion coefficients in the $x$
(blue circles) and $y$ (red diamonds) as a function of energy in
log-log scale. Points are numerical results, lines are fits to
power laws $D_\xi(E)=D_\xi^0 (E/\bar{V})^{\gamma_\xi}$, $\xi=x,y$.
b: Energy distribution for the experimental parameters. The shaded
regions correspond to sub-diffusive regimes (see text).}
\label{FigDiff}
\end{figure}
In order to understand our experimental findings in more details,
we have performed numerical simulations. In the experiment,
$k_{\textrm{B}} T \approx 6 \hbar \omega$, so that a few vertical
harmonic oscillator states are populated. The vertical size of the
atomic cloud, $\Delta z \simeq \sqrt{k_{\textrm{B}} T/m\omega}
\approx 1\mu$m (where $m$ is the atom mass), is much smaller than
$\sigma_\textrm{long}$ and the speckle potential can be considered
invariant along its propagation axis. Since it makes an angle
$\theta=30^{\circ}$ with respect to the expansion plane, it
couples the vertical quantum states. To account for these features
in the numerics, we consider the 3D dynamics of classical
particles in the external potential $V(x,y,z)+m \omega^2 z^2/2$,
with $V(x,y,z)=V_{\mathrm{iso}}(x\sin\theta - z\cos\theta,y)$
where $V_{\mathrm{iso}}(u, v)$ is a 2D \textit{isotropic} speckle
potential with correlation length $\sigma_y$. A classical particle
model is reasonable since $k_\textrm{B}T/\hbar\omega \approx6$ and
$k \sigma_y\approx 5$ where $k=\sqrt{m k_\textrm{B}T}/\hbar$.

A characteristic time scale for the dynamics is the Boltzmann time
$\tau_\textrm{B}$, \textit{i.e.}\ the time after which the memory
of the direction is lost, which depends on the particle energy.
For long times ($t \gg \tau_\textrm{B}$), scattering from the
angled speckle potential redistributes the kinetic and potential
energies in 3D, so that the dynamics in the horizontal plane is
expected to depend on the 3D particle energy $E$. We hence
calculate the spatial variances $\langle \xi^2(E,t)\rangle$ as a
function of time, where $\xi=x,y$ and brackets indicate averaging
over disorder and over initial conditions corresponding to the
energy $E$. The evolution is described by $\langle
\xi^2(E,t)\rangle \simeq 2 D_\xi(E) t^{\gamma_\xi(E)}$. We
identify three regimes characterized by the value of
$\gamma_\xi(E)$. For $E/\bar{V} \lesssim 2$, we find a
subdiffusive dynamics, \textit{i.e.}\ $\gamma_{\xi} (E) < 1$, for
experimentally-relevant time scales. In particular, for $E/\bar{V}
\lesssim 0.52$, we find strictly bounded trajectories,
$\gamma_{\xi} (E) = 0$. This is consistent with the percolation
threshold expected for 2D speckle potentials
\cite{Smith79,Weinrib82}. For $E/\bar{V} \gtrsim 2$, numerical
simulations yield a diffusive dynamics, \textit{i.e.}\
$\gamma_\xi(E) \simeq 1$. In this regime, the diffusion
coefficients $D_{\xi}(E)$ are strongly anisotropic and grow
algebraically with the particle energy (Fig.\,\ref{FigDiff}a).
From a fit to the numerical calculations for our parameters, we
find $D_x(E)=2.4 \sqrt{\bar{V}\sigma_y^2/m}(E/\bar{V})^{2.8}$ and
$D_y(E)=0.65 \sqrt{\bar{V}\sigma_y^2/m} (E/\bar{V})^{2.8}$
\cite{note, nfactor}. We have also done simulations of a classical
2D diffusion in the same anisotropic disorder. The various regimes
found in the 3D simulations with vertical confinement are also
found in 2D simulations at the same values of $E/\bar{V}$. In the
diffusive regime, the energy dependence of the diffusion
coefficients remains algebraic but with modified constants.

\begin{figure}[t!]
\centering
\includegraphics[width=0.45\textwidth]{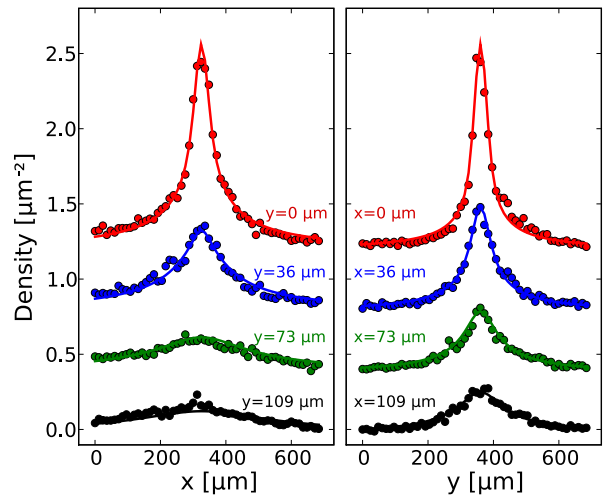}
\caption{Two-dimensional density distribution after 200\,ms of
expansion. Density profiles through cuts along $x$ (resp. $y$) at
different positions along $y$ (resp. $x$). The upper (red) points
are cuts through the central peak, whereas the other curves
downwards corresponds to positions separated by $\sim$36\,$\mu$m.
They are artificially offset for clarity. The lines are the result
from the 2D fit with Eq.\,(1) convolved by the imaging resolution
($\sim 15\,\mu$m).} \label{2Dcuts}
\end{figure}
In the experiment, the observed expansion results from the
diffusion of atoms with a broad energy distribution, $N(E)$. It
can be calculated assuming that, before abrupt release in the
horizontal plane, the gas is at thermal equilibrium in the trap
plus speckle potential. The corresponding energy distribution
$N(E) \propto \textrm{e}^{-E/k_\textrm{B} T} \sum_{n=0}^{E/\hbar
\omega} \left( 1 - \textrm{e}^{(n \hbar \omega -E)/ \bar V}
\right)$ is plotted in Fig.\,\ref{FigDiff}b \cite{formula}. It is
fully determined from the experimental parameters $\omega$, $N$,
$\bar{V}$ and $T$. Here, only 6\,\% of the atoms are sub-diffusive
($E\leq 2 \bar{V}$, shaded regions in Fig.\,4b). Incorporating
their contribution to the diffusive regime is thus a small error,
and for long expansion time, the column density can be
approximated by
\begin{equation}
n(x,y,t) \simeq \int_0^\infty \textrm{d}E\ N(E)\ \frac{1}{t} \
\frac{\exp\left(-\frac{x^2}{4D_x(E)t}
-\frac{y^2}{4D_y(E)t}\right)}{4\pi \sqrt{D_x(E) D_y(E)}}.
\end{equation}

We fit Eq.\,(1) (convolved with our imaging resolution $\sim
15\,\mu$m) to the experimental 2D density distribution with
$D_{x}(E)=D^0_{x}(E/E_\textrm{R})^\alpha$ and
$D_{y}(E)=D^0_{y}(E/E_\textrm{R})^\alpha$, $D^0_{x}$, $D^0_{y}$,
and $\alpha$ as fitting parameters, and the recoil energy
$E_\textrm{R}=k_\textrm{B} \times 180\,$nK as energy scale. As can
be seen on Fig.\,\ref{2Dcuts}, the 2D fit function reproduces the
data both close to the central peak and in the wings. We find
$D^0_{x}=3.0(1.5)\times 10^{-7}$m$^2$.s$^{-1}$,
$D^0_{y}=8.7(4.3)\times 10^{-8}$m$^2$.ms$^{-1}$, and
$\alpha=3.3(3)$.  The uncertainties come from the uncertainties on
the measurements of $N$, $\bar{V}$ and $T$ used in $N(E)$ and from
an observed systematic drift of the results as a function of the
expansion time \cite{ballistic}. Experimentally, the power-law
exponent is found to be $\alpha=3.3(3)$, to be compared with 2.8
in the simulation. The observed ratio of the two diffusion
coefficients is 3.45(15) when it is 3.7 in the simulation. These
slight discrepancies can be due to the approximations made in
order to derive Eq.\,(1). At $E=E_\textrm{R}\approx k_\textrm{B}
T$, numerically, we find $D_x(E_\textrm{R})=1.3(0.6)\times
10^{-7}$m$^2$.s$^{-1}$ and $D_y(E_\textrm{R})=3.5(1.7) \times
10^{-8}$m$^2$.s$^{-1}$, where the uncertainties come from the
uncertainties on the experimental parameters $\bar{V}$ and
$\sigma_y$. The experimental values of the diffusion coefficients
are thus in quantitative agreement with the 3D classical
simulation considering our relatively large uncertainties.

In conclusion, we have observed and studied 2D diffusive expansion
of ultra-cold atoms in a disordered potential. As a result of the
effective anisotropy of the speckle potential, the diffusion is
anisotropic. Fitting a diffusive model to our density profiles, we
are able to extract the diffusion coefficients and find a strong
dependence on the atom energy, in quantitative agreement with a
classical simulation for our parameters.

Understanding the diffusion properties as a function of energy (in
particular out of the weak scattering regime) is a necessary step
towards the study of other disorder-induced effects in two
dimensions, starting with anomalous subdiffusion \cite{Metzler00}
and classical trapping under the percolation threshold
\cite{Smith79, Weinrib82}. By cooling the gas further or by
reducing the correlation length of the disorder, we expect quantum
corrections to the diffusion and Anderson localization to show up
at the sub-mm length scale of the experiment \cite{Kuhn07}.
Moreover, in a 2D degenerate gas, the influence of disorder on the
Berezinskii-Kosterlitz-Thouless transition \cite{Dalibard06} is
especially intriguing. Will the vortices be pinned by disorder
\cite{Tung06}?

We thank F. Moron and A. Villing for technical assistance, M.
Besbes and GMPCS high performance computing facilities of the
LUMAT federation for numerical support. This research was
supported by CNRS, CNES as part of the ICE project, Direction
G\'en\'erale de l'Armement, ANR-08-blan-0016-01, IXSEA, EuroQuasar
program of the EU, and MAP program SAI of the European Space
Agency (ESA). LCFIO is member of IFRAF.


\begin{thebibliography}{0}

\bibitem[\dag]{email}{Corresponding author: thomas.bourdel@institutoptique.fr}

\bibitem[*]{presentaddress}{Present address: ETH Z\"urich, 8093 Z\"urich, Switzerland}

\bibitem{ashcroft1976}
N.W.~Ashcroft and N.D.~Mermin, \textit{Solid State Physics}
(Saunders, Philadelphia, USA, 1976).

\bibitem{degennes1966}
P.-G.~de~Gennes, {\it Superconductivity of Metals and Alloys}
(Benjamin, New York, 1966).

\bibitem{Mott68}
N. Mott, Rev. Mod. Phys. {\bf 40}, 677 (1968).

\bibitem{Anderson58}
P. Anderson, Phys. Rev. {\bf 109}, 1492(1958).

\bibitem{vollhardt1980}
D.~Vollhardt and P.~W\"olfle, Phys. Rev. Lett. \textbf{45}, 842
(1980).


\bibitem{Abrahams79} E. Abrahams, P.W. Anderson, D.C.
Licciardello, and T.V. Ramakrishnan, Phys. Rev. Lett. {\bf 42},
673 (1979).

\bibitem{Abrahams01}
E. Abrahams, S.V. Kravchenko, M.P. Sarachik, Rev. Mod. Phys. {\bf
73}, 251 (2001).

\bibitem{Kravchenko04}
S.V. Kravchenko and M.P. Sarachik, Rep. Prog. Phys. {\bf 67}, 1
(2004).

\bibitem{Allison06}
G. Allison {\it et al}, Phys. Rev. Lett. {\bf 96}, 216407 (2006).

\bibitem{Tracy09}
L. A. Tracy {\it et al}, Phys. Rev. B {\bf 79}, 235307 (2009).

\bibitem{Dubi07}
Y. Dubi, Y. Meir, and Y. Avishai, Nature {\bf 449}, 876 (2007).


\bibitem{Fallani08}
L. Fallani, C. Fort, and M. Inguscio, \textit{Adv. At. Mol. Opt.
Phys.} \textbf{56}, 119 (Academic Press, 2008).


\bibitem{Sanchez-Palencia09}
L. Sanchez-Palencia and M. Lewenstein, Nature Phys. \textbf{6}, 87
(2010).

\bibitem{Billy08}
J. Billy {\it et al}, Nature {\bf 453}, 891 (2008).

\bibitem{Roati08}
G. Roati {\it et al}, Nature {\bf 453}, 895 (2008).

\bibitem{Deissler09}
B. Deissler {\it et al}, arXiv:0910.5062 (2009).

\bibitem{White09}
M. White {\it et al}, Phys. Rev. Lett. {\bf 102}, 055301 (2009).

\bibitem{Pasienski09}
M. Pasienski, D. McKay, M. White and B. DeMarco, arxiv:0908.1182
(2009).

\bibitem{Grynberg00}
G. Grynberg, P. Horak, and C. Mennerat-Robilliard, Europhys. Lett.
\textbf{49}, 424 (2000).

\bibitem{Stern09}
G. Stern et al., arXiv:1003.4761

\bibitem{Meyrath05}
T. P. Meyrath, F. Schreck, J. L. Hanssen, C. -S. Chuu, and M. G. Raizen, Optics Express {\bf 13}, 2843 (2005).


\bibitem{Smith05}
N. L. Smith, W. H. Heathcote, G. Hechenblaikner, E. Nugent and C. J. Foot, J. Phys. B {\bf 38}, 223 (2005).

\bibitem{Goodman}
J.W. Goodman, \textit{Speckle Phenomena in Optics} (Roberts, Greenwood
Village, Colorado, 2007).


\bibitem{Clement06}
D. Cl\'ement {\it et al}, New J. Phys. {\bf 8}, 165 (2006).


\bibitem{Clement09}
J.-F. Cl\'ement {\it et al}, Phys. Rev. A {\bf 79}, 061406(R)
(2009).


\bibitem{Shapiro07}
B. Shapiro, Phys. Rev. Lett. {\bf 99}, 060602 (2007).


\bibitem{Smith79}
L. N. Smith and C. J. Lobb, Phys. Rev. B {\bf 20}, 3653 (1979).


\bibitem{Weinrib82}
A. Weinrib, Phys. Rev. B {\bf 26}, 1352 (1982).


\bibitem{note}
An algebraic increase of the diffusion coefficients with the
energy is also found in the weak scattering limit for isotropic
speckle potentials, with $\gamma= 2.5$ \cite{Kuhn07}. However, the
experiment is not in this regime.

\bibitem{nfactor}
The numerical factors depend on $\hbar \omega/\bar{V}$ and
$\sigma_x/\sigma_y$.

\bibitem{formula}
The vertical harmonic oscillator is quantized and the zero point
energy is taken as the energy of the ground state.

\bibitem{ballistic}
For atoms with energy $k_\textrm{B} T$, we find
$\tau_\textrm{B}\sim 50\,$ms and these atoms behave almost
diffusively after 200\,ms. However, some atoms with larger energy
remain ballistic over the time scale of the experiment. These
atoms could be responsible for the observed drift as the function
of expansion time.

\bibitem{Metzler00}
R. Metzler and J. Klafter, Phys. Rep. {\bf 339}, 1 (2000).

\bibitem{Kuhn07}
R. C. Kuhn, O. Sigwarth, C. Miniatura, D. Delande, and C. A.
M\"uller, New J. Phys. {\bf 9}, 161 (2007).

\bibitem{Dalibard06}
Z. Hadzibabic, P. Kr\"uger, M. Cheneau, B. Battelier  and J. Dalibard, Nature {\bf 441}, 1118 (2006).


\bibitem{Tung06}
S. Tung, V. Schweikhard and E. A. Cornell, Phys. Rev. Lett. {\bf
97}, 240402 (2006).


\end{thebibliography}
\end{document}